\documentclass[twocolumn]{aastex631}

\usepackage[version=4]{mhchem}
\usepackage{array}
\usepackage{lipsum}
\usepackage{soul}
\usepackage{wrapfig}
\usepackage{multirow}
\usepackage{amsmath}
\usepackage{hyperref}
\usepackage{textcomp}
\usepackage{gensymb}

\makeatletter
\DeclareRobustCommand{\HII}{%
  \mbox{H\check@mathfonts\fontsize\sf@size\z@\selectfont II}%
}
\makeatother

\begin{document}

\title{Mapping Physical Conditions in Neighboring Hot Cores: NOEMA Studies of W3(\ce{H_{2}O}) and W3(OH)}

\author{Morgan M. Giese}
\affiliation{Department of Astronomy, University of Wisconsin-Madison \\
475 N Charter St, Madison, Wisconsin 53706, USA}

\author{Will E. Thompson}
\affiliation{Department of Chemistry, University of Wisconsin-Madison \\
1101 University Ave, Madison, Wisconsin 53706, USA}

\author{Dariusz C. Lis}
\affiliation{Jet Propulsion Laboratory, California Institute of Technology \\
4800 Oak Grove Drive, Pasadena, CA, 91109, USA}

\author{Susanna L. Widicus Weaver}
\affiliation{Department of Astronomy, University of Wisconsin-Madison \\
475 N Charter St, Madison, Wisconsin 53706, USA}
\affiliation{Department of Chemistry, University of Wisconsin-Madison \\
1101 University Ave, Madison, Wisconsin 53706, USA}

\begin{abstract}

The complex chemistry that occurs in star-forming regions can provide insight into the formation of prebiotic molecules at various evolutionary stages of star formation. To study this process, we present millimeter-wave interferometric observations of the neighboring hot cores W3(\ce{H2O}) and W3(OH) carried out using the NOEMA interferometer. We have analyzed distributions of six molecules that account for most observed lines across both cores and have constructed physical parameter maps for rotational temperature, column density, and velocity field with corresponding uncertainties. We discuss the derived spatial distributions of these parameters in the context of the physical structure of the source. We propose the use of \ce{HCOOCH3} as a new temperature tracer in W3(\ce{H2O}) and W3(OH) in addition to the more commonly used \ce{CH3CN}. By analyzing the physically-derived parameters for each molecule across both W3(\ce{H2O}) and W3(OH), the work presented herein further demonstrates the impact of physical environment on hot cores at different evolutionary stages.
\end{abstract}

\keywords{Astrochemistry (75) --- Star-forming regions (1565) --- Interferometry (808) --- {\HII}
 regions (694)}

\section{Introduction} \label{sec:intro}

Complex chemistry occurs at every stage in the star formation process, but significant enhancement of complexity occurs after the collapse of a molecular cloud \citep{Blake1998, VanDishoeck2006, Herbst2009}. In this phase, the icy mantles of dust grains undergo radiative processing via cosmic rays and ultraviolet radiation to form complex organics and other prebiotic molecules \citep{MunozCaro2002, Oberg2011, Altwegg2019}. These prebiotic molecules are commonly studied in star-forming regions and comets, as many of these molecules make up a large majority of the chemical inventory in protoplanetary disks and are crucial in the development of planetary systems \citep{Blake1998, Jorgensen2012, WidicusWeaver2012, Altwegg2019, Rivilla2020, Ligterink2022}. While it is unclear as to how these molecules survive and are delivered to early-stage planets, they could possibly be brought to the system via meteorites and cometary impacts \citep{Oro1961, Chyba1990, Chyba1992, Blake1998, VanDishoeck2006, Hartogh2011}. Rotational transitions of prebiotic molecules are further useful for studying different stages of star formation as they can provide important information on physical characteristics such as temperatures and densities \citep{Tycho2021}. In order to probe these regions, the use of millimeter-wave interferometers has been instrumental in identifying how the physical environment of an interstellar region affects the chemistry that occurs \citep{Jorgensen2004, Tobin2011, Tycho2021}.

The neighboring hot cores W3(\ce{H2O}) and W3(OH) provide a unique environment to study how the molecular complexity can differ across different sources. The two cores are at different stages of stellar evolution, with W3(OH) being an ultra-compact {\HII} region \citep{Helmich1997, Rivera-Ingraham2013, Qin2015, WidicusWeaver2017, Thompson2023}.  Given that they formed from the same parent cloud, these two cores are well-suited for studying how prebiotic chemistry evolves during the star formation process. Both cores exhibit similar, yet distinct, molecular compositions, likely due to their difference in ages. W3(\ce{H2O}) and W3(OH) are such-named due to the presence of \ce{H2O} and OH masers, respectively, and are at a distance of $\sim$ 2 kpc from Earth with a v$_{lsr}$ of --47 km s$^{-1}$ \citep{Wynn1972, Wyrowski1997, Wilner1999, Rivera-Ingraham2013}. 

\cite{Thompson2023} recently presented results from millimeter-wave interferometric observations of these two cores. Building on this work, herein we present  further chemical analysis of the six molecules that were quantitatively analyzed in \cite{Thompson2023}.  This current study includes detailed parameter maps for these molecules in the two cores based on rotational temperature, column density, and velocity shift relative to v$_{lsr}$. Parameter maps show a visualization of the physical conditions within the region, allowing for a deeper insight into the effects of star formation on prebiotic molecules in hot cores.

\section{Observations and Data Reduction}

While specific details on the observations and data reduction process are described in \cite{Thompson2023}, an overview of the observations is described here. The two hot cores W3(\ce{H2O}) and W3(OH) were observed in 2021 using the IRAM/NOEMA interferometer\footnote{IRAM is supported by INSU/CNRS (France), MPG (Germany), and IGN (Spain).} in the C and D configurations with frequency coverage from 127.823 - 135.311 GHz and 143.116 - 150.666 GHz across the two receiver sidebands. Within the two sidebands, 28 high-resolution spectral windows were selected with channel spacings of 62.5 kHz to better identify and image molecular lines. The data sets were then reduced using the GILDAS packages CLIC and MAPPING\footnote{http://www.iram.fr/IRAMFR/GILDAS}, resulting in a synthesized beam of $\sim$ 1.87\arcsec ~$\times$ 1.34\arcsec ~(PA = --9.3\degree) in the lower sideband and $\sim$ 1.69\arcsec ~$\times$ 1.22\arcsec ~(PA = --7.9\degree) in the upper sideband. Figure 1 in \cite{Thompson2023} shows the imaged continuum at 132 GHz. At this frequency, the continuum emission from W3(OH) consists mostly of free-free emission, while that of W3(\ce{H2O}) is due to dust \citep{Wilner1995,Wyrowski1997,Wyrowski1999,Stecklum2002}.

The spectral fitting procedure described in this work utilizes the Global Optimization and Broadband Analysis Software for Interstellar Chemistry (GOBASIC) in a manner similar to that presented in previous works  \citep{WidicusWeaver2017, Zou2017, Wright2022, Thompson2023}. Specifics on GOBASIC operation are described in \cite{Rad2016}, with additional modifications described in \cite{Thompson2023}. Furthermore, the fitting procedure used in this work directly parallels the procedure described in-depth in \cite{Thompson2023}. However, a very brief discussion of GOBASIC is provided here. Rather than fitting individual lines in the spectrum for each molecule, GOBASIC performs a broadband analysis of molecular spectra by comparing the observational spectra to molecular catalog information provided by the Cologne Database for Molecular Spectroscopy and the Jet Propulsion Laboratory (JPL) Spectral Line Catalog \citep{Muller2001,Muller2005,Picket1998}. By assuming local thermodynamic equilibrium (LTE) and Gaussian line shapes, GOBASIC utilizes the Pattern Search Algorithm to iteratively vary the physical parameters until the simulation matches the observations and a global minimum fit is found.  GOBASIC is useful in that it can fit multiple molecules and physical components simultaneously, allowing for a thorough and comprehensive analysis of the spectra. Through this analysis, values for column density in cm$^{-2}$, spectral line full width half maximum ($\Delta v$) in km s$^{-1}$, rotational temperature in K, and velocity shift relative to the v$_{lsr}$  in km s$^{-1}$ can be derived. In this work, molecular spectra were first extracted at each pixel from the data cubes. Each pixel is smaller than the synthesized beam at $\sim$ 0.23\arcsec ~$\times$ 0.23\arcsec. They were then analyzed using GOBASIC to obtain parameter values for the v = 0 states of \ce{SO2}, \ce{CH3CN}, \ce{CH3CH2CN}, and \ce{CH3OCH3}, the v$_{t}$ = 0--2 state of \ce{CH3OH}, and the v$_{t}$ = 0--1 state for \ce{HCOOCH3}.  These components account for the majority of detected spectral lines in the observed frequency ranges. The spectrum extracted at the pixel corresponding to the peak of the continuum flux in each core can be seen in \cite{Thompson2023}.

\section{Parameter Maps} \label{sec:maps}

\begin{figure*}[t!]
  \includegraphics[width=\textwidth]{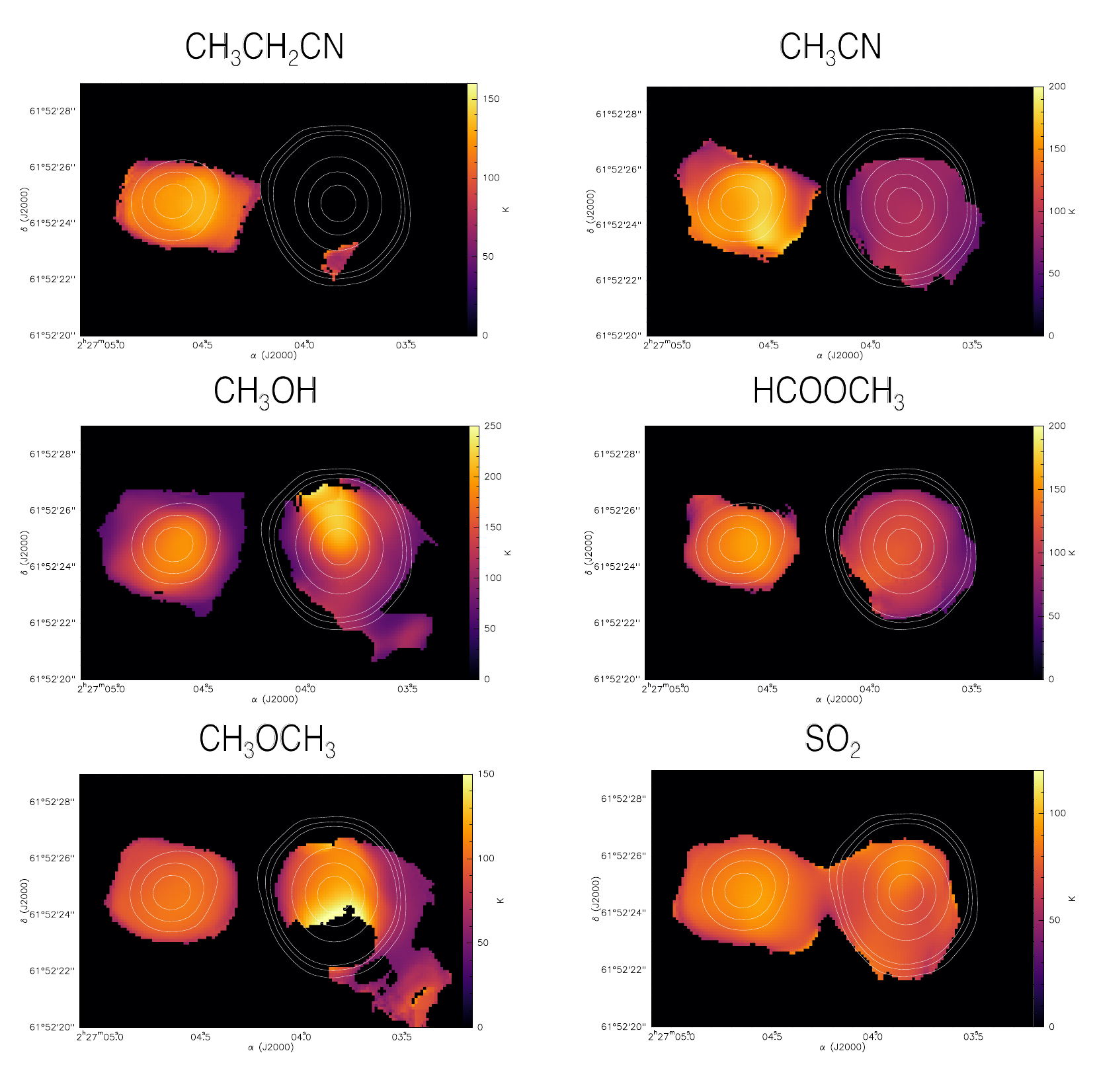}
  \caption{Temperature maps for the six molecules obtained by fitting the spectrum at each individual pixel. All six molecules were fit simultaneously using GOBASIC. The white contour levels correspond to the continuum levels in \cite{Thompson2023} at 12, 18, 25, 100, 200, 300 times $\sigma$ ($\sigma$ = 6.481 mJy beam$^{-1}$).
  \label{fig:temp_maps}}
\end{figure*}

As demonstrated in \cite{Thompson2023}, the NOEMA spectral cubes can be used to create parameter maps for the detected molecules that have a sufficient number of lines to enable robust spectral fitting. General specifics on the process of creating these maps are described in \cite{Thompson2023}, though the current work includes slight modifications. From the spectra, fits of all six molecules were conducted using GOBASIC to derive values for column density, temperature, and velocity shift relative to the v$_{lsr}$. Maps were then constructed using this information via the methods described in \cite{Thompson2023}. Previously, only temperature and column density maps were created for \ce{CH3OH} by analyzing it as a singular molecular component with GOBASIC. Here, all six molecules were fit simultaneously using GOBASIC. From this, updated temperature, column density, and velocity shift maps were determined for \ce{CH3OH}, along with new maps for \ce{CH3CH2CN}, \ce{CH3CN}, \ce{CH3OCH3}, \ce{HCOOCH3}, and \ce{SO2}.  The temperature, column density, and velocity shift maps for each molecule are shown in Figures \ref{fig:temp_maps}, \ref{fig:dens_maps}, and \ref{fig:vel_maps}, respectively. Uncertainty maps for each parameter can be found in Appendix \ref{sec:uncertain}.  As \ce{CH3OH}, \ce{CH3CN}, \ce{CH3OCH3}, and \ce{HCOOCH3} have many strong spectral lines in the observed frequency range, they are good choices for this mapping process.  Likewise, these molecules have been used in the past to study this source \citep{Ahmadi2018, Qin2015, Chen2006, WidicusWeaver2017, Thompson2023}. In addition, \ce{CH3CH2CN} was included in the current analysis because it was previously well-fit in W3(\ce{H2O}) \citep{WidicusWeaver2017, Thompson2023}. While \ce{CH3CH2CN} was not previously well-fit in W3(OH), the results presented here show that  some pixels near the southern side of the core can be fit with accurate values. \ce{SO2} was also included in this analysis because it was well-fit across both cores and is prominent in ultra-compact {\HII} regions \citep{Minh2010}. For both the temperature and column density maps, there is a slight offset between the peak values and the peak of the continuum emission. This is further discussed in Section \ref{sec:shift}. 

\begin{figure*}[t!]
  \includegraphics[width=\textwidth]{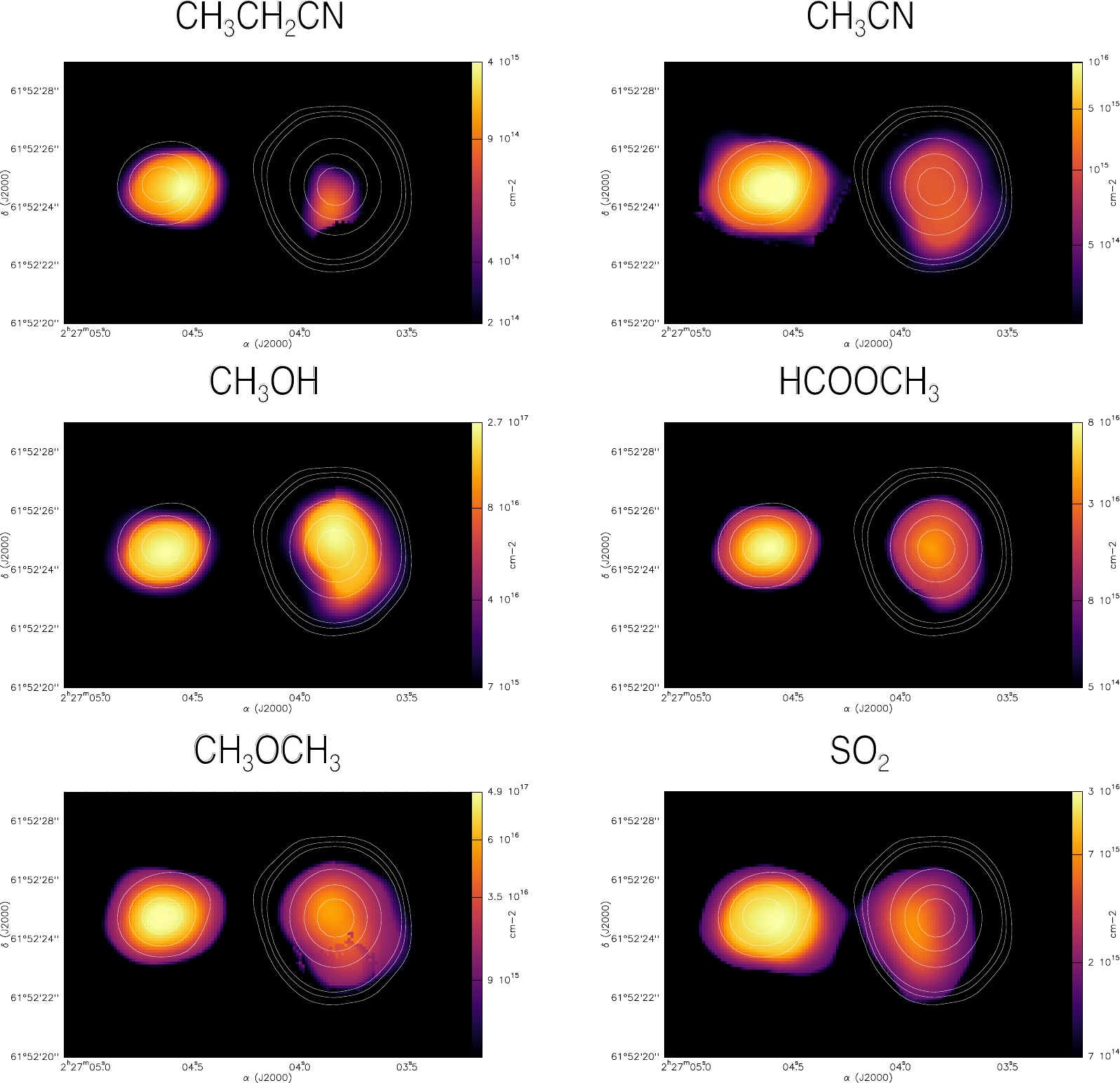}
  \caption{Column density maps for the six molecules obtained by fitting the spectrum at each individual pixel. All six molecules were fit simultaneously using GOBASIC. The white contour levels correspond to the continuum levels in \cite{Thompson2023} at 12, 18, 25, 100, 200, 300 times $\sigma$ ($\sigma$ = 6.481 mJy beam$^{-1}$).
  \label{fig:dens_maps}}
\end{figure*}

\begin{figure*}[t!]
  \includegraphics[width=\textwidth]{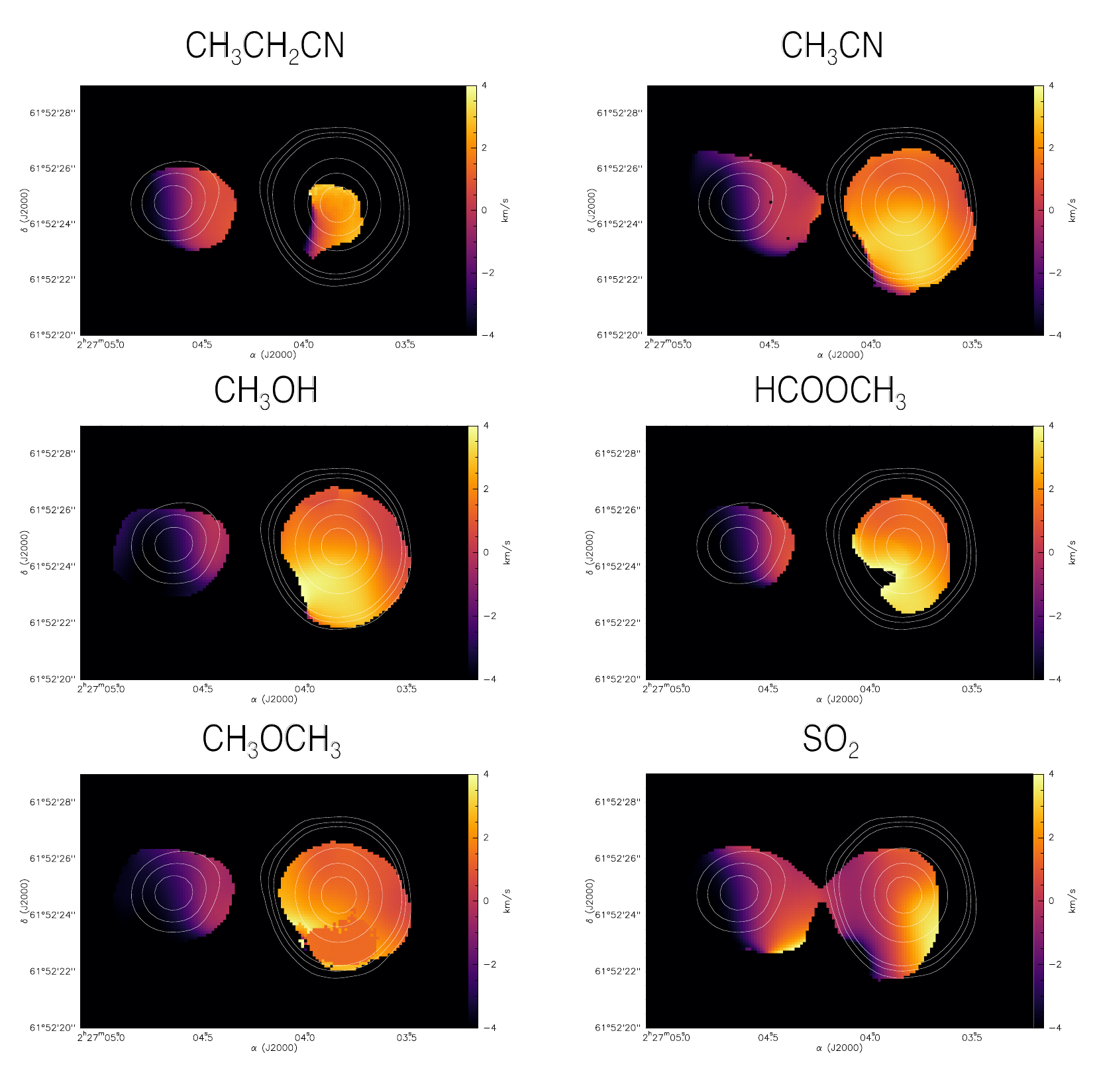}
  \caption{Velocity shift maps for the six molecules obtained by fitting the spectrum at each individual pixel. All six molecules were fit simultaneously using GOBASIC. The white contour levels correspond to the continuum levels in \cite{Thompson2023} at 12, 18, 25, 100, 200, 300 times $\sigma$ ($\sigma$ = 6.481 mJy beam$^{-1}$).
  \label{fig:vel_maps}}
\end{figure*}

\subsection{CH\texorpdfstring{$_3$}{3}CH\texorpdfstring{$_2$}{2}CN}

\ce{CH3CH2CN} proved difficult to fit in W3(OH) due to severe line blending from unidentified molecules in the spectra as described by \cite{Thompson2023}. Nevertheless, spectra from a small portion of the core had a sufficient number of unblended lines to enable determination of parameters. The maps show both a higher temperature and a higher column density for \ce{CH3CH2CN} in W3(\ce{H2O}) as compared to W3(OH). \ce{CH3CH2CN} in W3(\ce{H2O}) peaks around 130 K and 4.0$\times$10$^{15}$ cm$^{-2}$ with a gradual increase in velocity shift from around --4.0 km s$^{-1}$ to 1.0 km s$^{-1}$ east to west. The few well-fit pixels in W3(OH) peak at lower values around 100 K and 9.0$\times$10$^{14}$ cm$^{-2}$ with a gradual increase in velocity shift from around --2.0 km s$^{-1}$ to 3.0 km s$^{-1}$ east to west. This analysis not only confirms the presence of \ce{CH3CH2CN} in W3(OH), but also provides the first quantitative analysis of the molecule in this region.

\subsection{CH\texorpdfstring{$_3$}{3}CN} \label{sec:ch3cn}

Similar to the procedures followed by \cite{Ahmadi2018} and \cite{Thompson2023}, precautions were taken when fitting parameters for \ce{CH3CN}. For each spectra across both cores, the 8(7) -- 7(7), 8(6) -- 7(6), and 8(5) -- 7(5) transitions were excluded from the fit because they were severely blended with other molecules. \ce{CH3CN} shows both a higher temperature and a higher column density in W3(\ce{H2O}). We find that \ce{CH3CN} peaks in W3(\ce{H2O})  around 190 K and 1.0$\times$10$^{16}$ cm$^{-2}$, with a gradual increase in velocity shift from around --4.0 km s$^{-1}$ to 1.0 km s$^{-1}$ east to west. \ce{CH3CN} in W3(OH) peaks around 100 K and 3.0$\times$10$^{15}$ cm$^{-2}$ with a gradual increase in velocity shift from around 1.5 km s$^{-1}$ to 3.5 km s$^{-1}$ north to south before decreasing down to --2.0 km s$^{-1}$ at the southern-most region.

\subsection{CH\texorpdfstring{$_3$}{3}OCH\texorpdfstring{$_3$}{3}}

\ce{CH3OCH3} shows a higher temperature in W3(OH), yet a higher column density in W3(\ce{H2O}). We find that \ce{CH3OCH3} peaks in W3(\ce{H2O}) around 120 K and 4.8$\times$10$^{17}$ cm$^{-2}$, with a gradual increase in velocity shift from around --4.0 km s$^{-1}$ to 0.5 km s$^{-1}$ east to west. \ce{CH3OCH3} in W3(OH) peaks around 150 K and 6.0$\times$10$^{16}$ cm$^{-2}$, with a gradual decrease in velocity shift from around 4.0 km s$^{-1}$ to 1.0 km s$^{-1}$ east to west. The temperature map for \ce{CH3OCH3} presents an intriguing question: what is causing the weak spectra for \ce{CH3OCH3} on the southeast side of W3(OH) as compared to the rest of the core? In the column density and velocity shift maps, this region is characterized by a stark decrease in value compared to those derived from the surrounding spectra. \cite{Ahmadi2018} sees two intensity peaks in W3(OH) in the integrated intensity map for \ce{CH3CN} around $\sim$ $\delta$(J2000) = 61$^{o}$52\arcmin25\arcsec.5 and $\sim$ $\delta$(J2000) = 61$^{o}$52\arcmin23\arcsec, with the latter peak being consistent with the area of interest in the current work. This dramatic change in physical parameters as compared to the rest of the core could be indicative of a distinct physical component in the core structure. Further analysis with higher spatial resolution could be beneficial to solving this mystery.

\subsection{CH\texorpdfstring{$_3$}{3}OH}

As mentioned above, new \ce{CH3OH} maps were created from fits utilizing multiple molecular components. In these maps, \ce{CH3OH} shows a higher temperature in W3(OH) and similar column densities in both cores.  \ce{CH3OH} peaks in W3(\ce{H2O}) around 200 K and 2.7$\times$10$^{17}$ cm$^{-2}$, with a gradual increase in velocity shift from around --4.0 km s$^{-1}$ to 0.0 km s$^{-1}$ east to west. \ce{CH3OH} peaks in W3(OH) around 220 K and 2.7$\times$10$^{17}$ cm$^{-2}$ with a gradual increase in velocity shift from around 1.0 km s$^{-1}$ to 3.5 km s$^{-1}$ north to south. The results for temperature and column density are consistent with the results presented by \cite{Thompson2023}.

\subsection{HCOOCH\texorpdfstring{$_3$}{3}}

\ce{HCOOCH3} shows both a higher temperature and a higher column density in W3(\ce{H2O}). \ce{HCOOCH3} peaks in W3(\ce{H2O}) around 150 K and 8.0$\times$10$^{16}$ cm$^{-2}$, with a gradual increase in velocity shift from around --4.0 km s$^{-1}$ to 1.0 km s$^{-1}$ east to west. \ce{HCOOCH3} peaks in W3(OH)  around 120 K and 3.0$\times$10$^{16}$ cm$^{-2}$, with a gradual increase in velocity shift from around 2.0 km s$^{-1}$ to 4.0 km s$^{-1}$ north to south. Interestingly, the column density and velocity shift maps for \ce{HCOOCH3} seem to be affected by a similar phenomenon as the maps for \ce{CH3OCH3}, although to a lesser degree. The decrease in column density in this area of W3(OH) is small, and the velocity shift is poorly determined in a small region on the southwest side.

\subsection{SO\texorpdfstring{$_2$}{2}}

\ce{SO2} shows similar temperatures in both cores and a higher column density in W3(\ce{H2O}). \ce{SO2} in W3(\ce{H2O}) peaks around 100 K and 3.0$\times$10$^{16}$ cm$^{-2}$, and in W3(OH) peaks around 100 K and 7.0$\times$10$^{15}$ cm$^{-2}$. The velocity shift map for \ce{SO2} is unlike that for any of the other molecules in this analysis. From east to west encompassing both cores, the velocity first increases from around --4.0 km s$^{-1}$ to 1.0 km s$^{-1}$, proceeds to decrease to --1.0 km s$^{-1}$, before finally increasing up to 4.0 km s$^{-1}$. The southwest region of W3(\ce{H2O}) and the southeast region of W3(OH) tend to the extreme with values around 4.0 km s$^{-1}$ and --4 km s$^{-1}$, respectively. All of the \ce{SO2} maps show a lack of molecular emission from the west side of W3(OH). This is consistent with the integrated intensity map shown in \cite{Thompson2023}.

Sulfur chemistry is widely used as an indicator of age in hot cores \citep{Helmich1994, Charnley1997, Hatchell1998, Wakelam2004}. Previously created models for young hot cores predict \ce{SO2} to first increase, then decrease in abundance as the core progresses in its evolutionary cycle \citep{Millar1997, Hatchell1998}. This eventual decrease in abundance is evident in the column density map for \ce{SO2} with peak values for W3(\ce{H2O}) being higher than those for W3(OH), as W3(OH) is at a later evolutionary stage than W3(\ce{H2O}). \ce{SO2} is the only molecule shown with a temperature map that does not have significant differences in temperature between W3(\ce{H2O}) and W3(OH), despite the differences in column density between the cores.

\section{Discussion}

The parameter maps for the W3 cores that are presented above reveal new physical insight into the chemistry, structure, and kinematics of these sources. These findings are discussed below.

\subsection{Shift in Peak Emission Position in W3(H\texorpdfstring{$_2$}{2}O)} \label{sec:shift}

The temperature and density peaks for \ce{CH3CH2CN}, \ce{CH3CN}, \ce{HCOOCH3}, and \ce{SO2} in W3(\ce{H2O}) are slightly offset to the west from the peak of the continuum emission. This can be attributed to the presence of two embedded protostars, named W3(\ce{H2O}) E and W3(\ce{H2O}) W, within the core \citep{Wyrowski1999, Ahmadi2018}. W3(\ce{H2O}) E is located at $\alpha$(J2000) = 02$^{h}$27$^{m}$04.73$^{s}$, $\delta$(J2000) = 61$^{o}$52\arcmin24\arcsec.66 and W3(\ce{H2O}) W at $\alpha$(J2000) = 02$^{h}$27$^{m}$04.57$^{s}$, $\delta$(J2000) = 61$^{o}$52\arcmin24\arcsec.59 \citep{Ahmadi2018}. Marked positions of the two protostars can be seen in \cite{Ahmadi2018}. The maps described in Section \ref{sec:maps} show that in W3(\ce{H2O}), the molecular emission traces W3(\ce{H2O}) W while the continuum emission traces W3(\ce{H2O}) E. This is consistent with the findings of \cite{Ahmadi2018} for dense gas tracers such as \ce{CH3CN} and is seen in their integrated intensity maps.

\subsection{Comparing Column Density and Temperature Maps}

The temperature and column density maps of the complex organic molecules (COMs) reveal that there is not a single value of temperature that can describe the detected molecules in a given core. Temperatures of COMs in both cores range from $\sim$50 K to more than 200 K. Likewise, as is expected, column densities vary greatly.  However, it appears that the COMs follow similar trends in terms of source kinematics.  All COMs display similar velocity shifts. In W3(\ce{H2O}), the velocity shift increases moving east to west. This is similar to the velocity gradient seen in \cite{Ahmadi2018} and provides further evidence for the division of W3(\ce{H2O}) into the two protostars W3(\ce{H2O}) W and W3(\ce{H2O}) E. In W3(OH), apart from the aforementioned southern region of \ce{CH3OCH3}, velocity shift increases moving north to south. This seems to be tracing the movement of the non-ionized molecular gas \citep{Keto1995, Wilner1999, Ahmadi2018}.

\begin{figure*}[t!]
  \includegraphics[width=\textwidth]{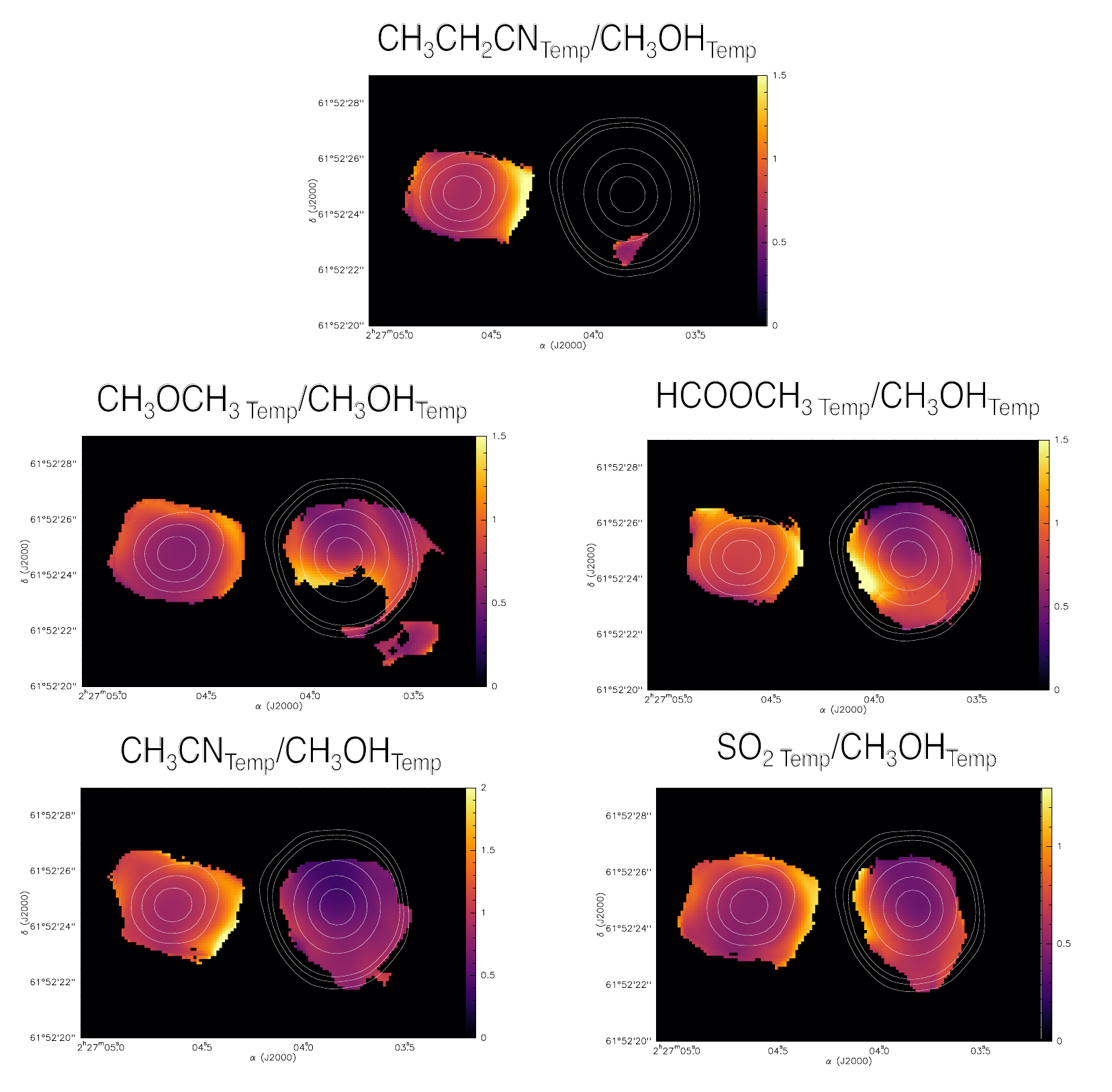}
  \caption{Temperature ratio maps comparing each molecule to \ce{CH3OH}. For each molecule, the derived temperature value at each pixel was divided by the corresponding \ce{CH3OH} temperature value. The white contour levels correspond to the continuum levels in \cite{Thompson2023} at 12, 18, 25, 100, 200, 300 times $\sigma$ ($\sigma$ = 6.481 mJy beam$^{-1}$).
  \label{fig:ratio_temp_maps}}
\end{figure*}

\begin{figure*}[t!]
  \centering
  \includegraphics[width=0.87\textwidth]{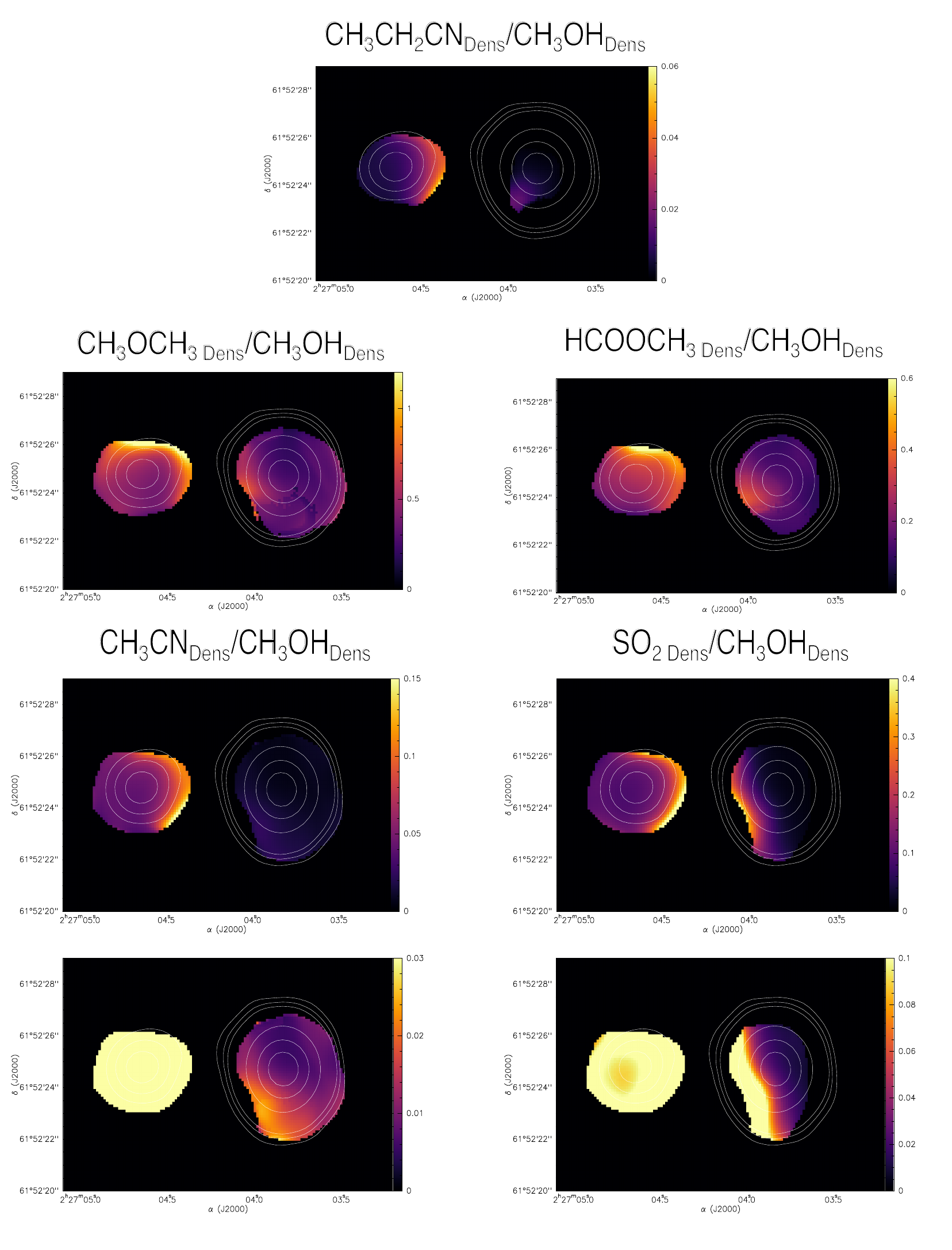}
  \caption{Column density ratio maps comparing each molecule to \ce{CH3OH}. For each molecule, the derived column density value at each pixel was divided by the corresponding \ce{CH3OH} column density value. The white contour levels correspond to the continuum levels in \cite{Thompson2023} at 12, 18, 25, 100, 200, 300 times $\sigma$ ($\sigma$ = 6.481 mJy beam$^{-1}$) The \ce{CH3CN} and \ce{SO2} ratio maps are shown using two maps to exhibit the vast difference between cores.
  \label{fig:ratio_dens_maps}}
\end{figure*}

In order to directly compare the temperature and column density values between different molecules, ratio maps between \ce{CH3OH} and the other molecules were created. \ce{CH3OH} was chosen to be the comparison molecule as it had the highest peak temperature and column density values out of the six well-fit molecules. The temperature and column density ratio maps are shown in Figures \ref{fig:ratio_temp_maps} and \ref{fig:ratio_dens_maps}.

The temperature ratio maps show that \ce{CH3OH} has a much steeper decrease in temperature values moving outward from the center of each core as compared to other molecules. The derived temperature values for \ce{CH3OH} are smaller than those of any of the other molecules on the edges of the cores in some locations. In the region of the cores near the continuum emission peak, \ce{CH3OH} is warmer than other molecules by up to a factor of two. The values for \ce{CH3CN} in W3(OH) are the outlier to this trend, with the calculated ratio staying within the range of 0.5 - 1.0. 

It is interesting to note the difference between the temperature ratio map of \ce{CH3OCH3}/\ce{CH3OH} and that of \ce{HCOOCH3}/\ce{CH3OH}. The \ce{CH3OCH3}/\ce{CH3OH} temperature ratio map has values $\sim$0.4 in the center of W3(\ce{H2O}), and the \ce{HCOOCH3}/\ce{CH3OH} temperature ratio map has values $\sim$0.75 towards the center of W3(\ce{H2O}). Similarly to that observed in the parameter maps, the region on the south side of W3(OH) that is showing unusual behavior is also evident in the temperature ratio maps for both \ce{CH3OCH3} and \ce{HCOOCH3}. In addition, this region is now apparent in the temperature ratio maps for \ce{CH3CN} and \ce{SO2}, albeit much less significant in these cases.

The column density ratio maps show a much more consistent ratio in both cores across all molecules than do the temperature maps, with most column density ratios peaking around 0.5. In W3(\ce{H2O}), the column density ratios tend to peak on the west side of the core, offset from the continuum peak. In W3(OH), column density ratios tend to peak on the east side of the core, again offset from the continuum peak. \ce{CH3OCH3} has the highest column density ratio to \ce{CH3OH} out of all of the molecules, peaking at over 1.0 on the edge of W3(\ce{H2O}). 
 
In W3(\ce{H2O}), both \ce{CH3OCH3} and \ce{HCOOCH3} have a ratio that is lower towards the center of the core and higher towards the outside of the core. Because optical depths are low for both molecules, this result is not due to any radiative transfer effect. These molecules are commonly thought of as tracers of grain surface chemistry, forming from the photolysis of \ce{CH3OH} on icy grains \citep{Garrod2008, Laas2011}. \ce{CH3OCH3} and \ce{HCOOCH3} would therefore be expected to have temperature and density profiles similar to that of \ce{CH3OH}. However, the column density ratio maps show this is not the case. The column density ratio map of \ce{CH3OCH3}/\ce{CH3OH} yields higher values outside of the continuum contours, indicating a higher column density of \ce{CH3OCH3} outside of the continuum, where there is less \ce{CH3OH}. It is not clear if this is an artifact of the spectral line fits for these molecules in the region outside of the continuum, or if there actually is a higher abundance of \ce{CH3OCH3} and \ce{HCOOCH3} in this region. When looking at the individual spectra in this region, there are no obvious indicators of artifacts. Furthermore, the signal-to-noise ratio is consistent with that in spectra from other areas in the core. If \ce{CH3OCH3} is forming from photolysis of \ce{CH3OH} in the ice, one would expect the density maps of \ce{CH3OCH3} and \ce{HCOOCH3} to trace the same spatial area as \ce{CH3OH}. The region of unusual structure once again appears in the maps for \ce{CH3OCH3}, \ce{HCOOCH3}, and \ce{CH3CN}.

\begin{figure*}[t!]
  \includegraphics[width=\textwidth]{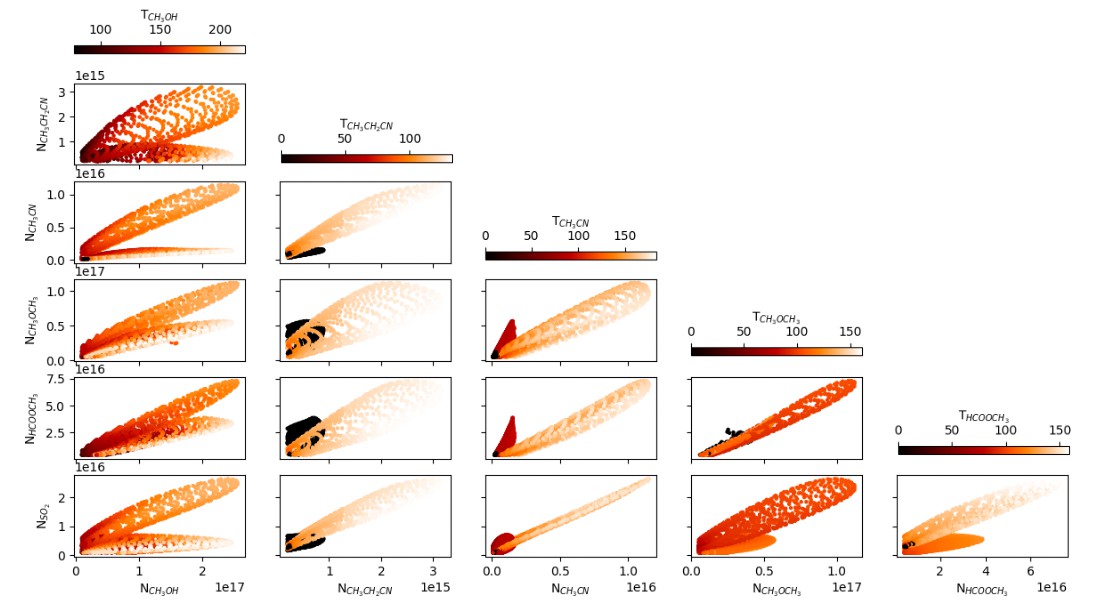}
  \caption{Plots comparing column densities in cm$^{-2}$ between two molecules. The color bar denotes temperature in K of the molecule on the x-axis. Black data points denote where the fits for column density do not overlap with the fits for temperature (most evident in the pixels in W3(OH) for \ce{CH3CH2CN}). It should be noted that since the pixel size is smaller than the synthesized beam, these plots are over-sampled. However, due to the asymmetric nature of the two cores, re-binning the data to account for the size of the synthesized beam would introduce additional averaging errors.
  \label{fig:coldens_comparison}}
\end{figure*}

\begin{figure*}[t!]
  \includegraphics[width=\textwidth]{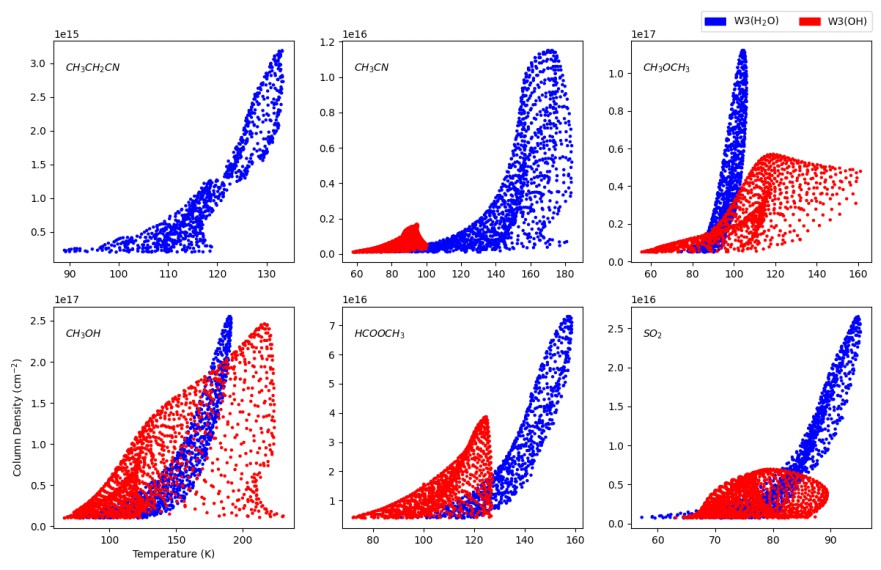}
  \caption{Plots comparing column density to temperature for each molecule. Derived parameters from spectra located in W3(\ce{H2O}) and W3(OH) are denoted by blue and red, respectively. \ce{CH3CH2CN} does not show results for W3(OH) as there were no regions with both well-derived column density and well-derived temperature.
  \label{fig:coldensvstemp}}
\end{figure*}

To further analyze the relationship in column densities and temperatures between each pair of molecules, we follow a similar technique used in \cite{Law2021} to create Figures \ref{fig:coldens_comparison} and \ref{fig:coldensvstemp}. 
In Figure \ref{fig:coldens_comparison}, we compare the calculated column density values at each individual pixel. Every plot has a clear separation into a component for W3(\ce{H2O}) and a component for W3(OH). Furthermore, most pairs of molecules have a direct correlation in column densities. The outlier to this trend lies in the comparison of molecules to \ce{CH3CH2CN} in W3(OH) due to the low signal-to-noise spectra in many regions of the core. However, the \ce{CH3CH2CN} values in W3(OH) correlate the best with those of \ce{CH3CN}, as a secondary component is seemingly beginning to form similar to that in the comparisons of other molecules. Further analysis of \ce{CH3CH2CN} in W3(OH) is needed to confirm this. \ce{CH3OH}, \ce{CH3OCH3}, and \ce{HCOOCH3} all exhibit similar spatial correlations in both cores in agreement with previous studies of these organics in star-forming regions \citep{Bisschop2007, WidicusWeaver2017, Law2021, Thompson2023}. The \ce{CH3CN} plots show interesting results similar to those described in \cite{Law2021}. While \ce{CH3CN} spatially correlates well with every molecule in W3(\ce{H2O}), it seems to only correlate well with \ce{CH3OH} in W3(OH). This is likely due to both molecules being related to total gas density as opposed to any relation in chemical formation pathways \citep{Garrod2006, Garrod2008, Law2021}.

Figure \ref{fig:coldensvstemp} compares column density and temperature values for each molecule. For all molecules except \ce{CH3CN}, W3(\ce{H2O}) exhibits a positive exponential trend. W3(OH) also exhibits a positive trend, however cannot be described with any specific function. This difference can most likely be attributed to the complex nature of the W3(OH). As W3(OH) is an ultra-compact {\HII} region, most of the gas originates from outside a dusty cocoon surrounding the embedded protostar, while the inner region is ionized \citep{Wynn1972, Qin2015, Thompson2023}. With two different sources of molecular gas in the core, it is difficult to ascertain a specific type of trend in W3(OH). \ce{CH3OH}, \ce{CH3OCH3}, and \ce{HCOOCH3} are the most obvious examples of this as both molecules show much less compact trends. As stated in \cite{Thompson2023}, these molecules are intrinsically linked via formation pathways as both \ce{CH3OCH3} and \ce{HCOOCH3} can form from chemistry involving methanol photodissociation and the \ce{CH3O} radical \citep{Garrod2006, Garrod2008, Laas2011, Garrod2022}.

\subsection{CH\texorpdfstring{$_3$}{3}CN as a Molecular Thermometer}

\ce{CH3CN} is commonly used as a molecular tracer for temperature in hot cores \citep{Loren1984, Zhang1998, Ahmadi2018}. It is a symmetric-top molecule with many transitions closely spaced, allowing for direct comparison between the excitation temperatures of similar transitions in different \textit{K}-ladders of the rotational spectrum \citep{Loren1984}. In using \ce{CH3CN} as a thermometer for hot cores, it is important to account for both non-LTE excitation and optical depth effects. As mentioned in Section \ref{sec:ch3cn}, the analysis presented here follows a similar method to \cite{Ahmadi2018} and \cite{Thompson2023}, specifically removing the $K$ = 5--7 lines from the analysis. Comparing the temperature map of W3(\ce{H2O}) for \ce{CH3CN} to that presented in \cite{Ahmadi2018}, the molecular emission seems to peak in similar locations across the core. However, this work shows peaks at a temperature roughly 100 K lower than the previous work.  The results presented here are more consistent with the calculated values for the other five molecules analyzed in this work. Even though similar methods were used in the two maps to account for difficult lines in the spectrum, wildly different results were achieved. One potential reason for the difference in results is that the analysis presented here better accounts for line blending in the \ce{CH3CN} transitions. However, the \ce{CH3CN} transitions presented in \cite{Ahmadi2018} do not seem to suffer from any blending issues apart from the CH$_3$$^{13}$CN isotopologue, which is accounted for in their fitting. It could also be due to the difference in rotational level energies, as the analyzed transitions presented here have lower energies than those in \cite{Ahmadi2018}.

To counter this problem in future studies of W3(\ce{H2O}) and W3(OH), we propose the use of \ce{HCOOCH3} as the main tracer for temperature in these specific star-forming cores. While \ce{HCOOCH3} has been used as a molecular tracer for temperature in similar sources, it has yet to be used for W3(\ce{H2O}) and W3(OH) \citep{Gorai2021, Brouillet2022}. \ce{HCOOCH3} is not a symmetric-top molecule like \ce{CH3CN} and as such does not have similar transitions in different \textit{K}-ladders for comparison. However, the use of \ce{HCOOCH3} as a tracer removes the need for accounting for optical depth effects. In all spectra across both cores, \ce{HCOOCH3} had no optically-thick transitions. Furthermore, when comparing the temperature maps of all six molecules, the values and emission shapes for \ce{CH3CN} are most similar to those for \ce{HCOOCH3} in each core.

\section{Conclusions}

We have carried out millimeter-wave observations of the W3(\ce{H2O}) and W3(OH) star-forming regions using the IRAM/NOEMA interferometer. From these observations, a spectrum at each individual pixel was extracted and fit using GOBASIC for the molecules \ce{CH3OH}, \ce{CH3CH2CN}, \ce{CH3CN}, \ce{CH3OCH3}, \ce{HCOOCH3}, and \ce{SO2}. Rotational temperature, column density, and velocity shift relative to the v$_{lsr}$ were then determined at each pixel.  Physical parameter maps for each of the six molecules were then created. Additionally, ratio maps were created for temperature and column density to compare each molecule to \ce{CH3OH}. Further plots were created to discuss the spatial comparison of these values between molecules.

Temperatures and column densities vary across the six molecules, ranging from 50 K to 200 K and 4.0$\times10^{14}$ cm$^{-2}$ to 4.9$\times10^{17}$ cm$^{-2}$, respectively. As \ce{CH3CN} has consistently proven difficult to accurately fit in studies of the two cores, we suggest \ce{HCOOCH3} as an alternative for tracing temperature in this region. \ce{SO2}, the only inorganic molecule analyzed in this study, revealed the most consistent temperature across both cores, peaking at $\sim$ 100 K in each core. \ce{SO2} also has a lower column density  in W3(OH), supporting previous models describing a decrease in in the molecule at later stages of star formation. Generally, the results from this study for temperature and column density are consistent with previous analyses of these molecules in this region \citep{Qin2015, WidicusWeaver2017, Ahmadi2018, Thompson2023}. Furthermore, the velocity shift maps support the claim made in \cite{Ahmadi2018} that W3(\ce{H2O}) is split into the two embedded protostars W3(\ce{H2O}) W and W3(\ce{H2O}) E. This claim is also supported by the shift in molecular peak emission to the west of the continuum peak emission that was observed in the current work.

When comparing the column densities of any pair of molecules, both cores exhibit a direct, positive correlation. While the N- and O-bearing organic species each follow their own similar trends, there is an overlap with \ce{CH3OH} and \ce{CH3CN} due to both being related to total gas density \citep{Garrod2006, Garrod2008, Law2021}. In W3(\ce{H2O}), each individual molecule apart from \ce{CH3CN} shows a compact, exponential trend when comparing column density and temperature. W3(OH) is more complicated because it is an {\HII} region, but generally shows a positive trend for each molecule.

\section{Acknowledgements} \label{sec:ack}
S.L.W.W., M.M.G., and W.E.T. thank the University of Wisconsin-Madison for S.L.W.W.'s startup support and access to NOEMA time that enabled this research.  We thank Ka Tat Wong from IRAM for support in setting up the observations and initial data reduction. Part of this research was carried out at the Jet Propulsion Laboratory, California Institute of Technology, under a contract with the National Aeronautics and Space Administration (80NM0018D0004).

\software{Astropy \citep{astropy:2013, astropy:2018, astropy:2022}}

\begin{appendix}

\section{Uncertainty Maps} \label{sec:uncertain}

\begin{figure*}[t!]
  \includegraphics[width=\textwidth]{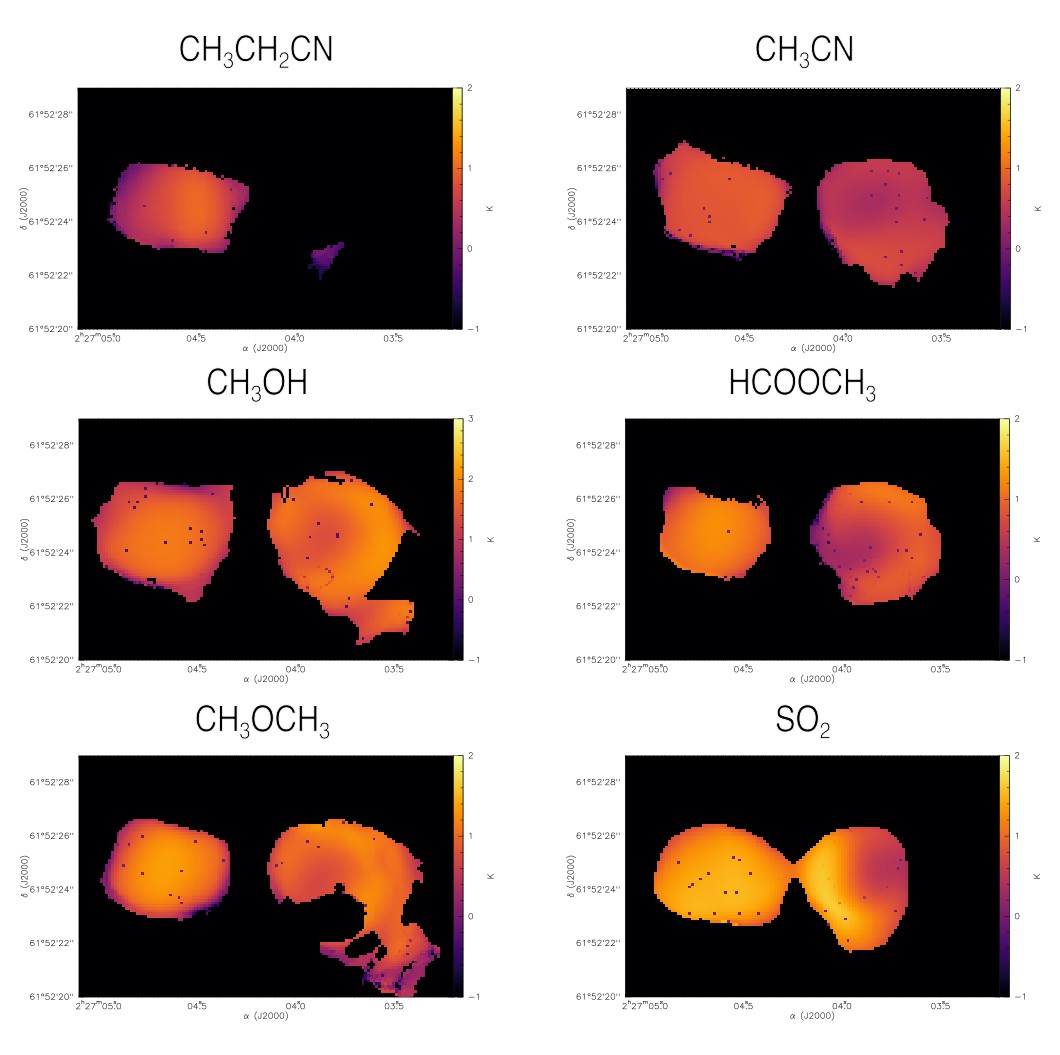}
  \caption{Maps describing the uncertainty of temperature values for each molecule.
  \label{fig:uncertaintytemp}}
\end{figure*}

\begin{figure*}[t!]
  \includegraphics[width=\textwidth]{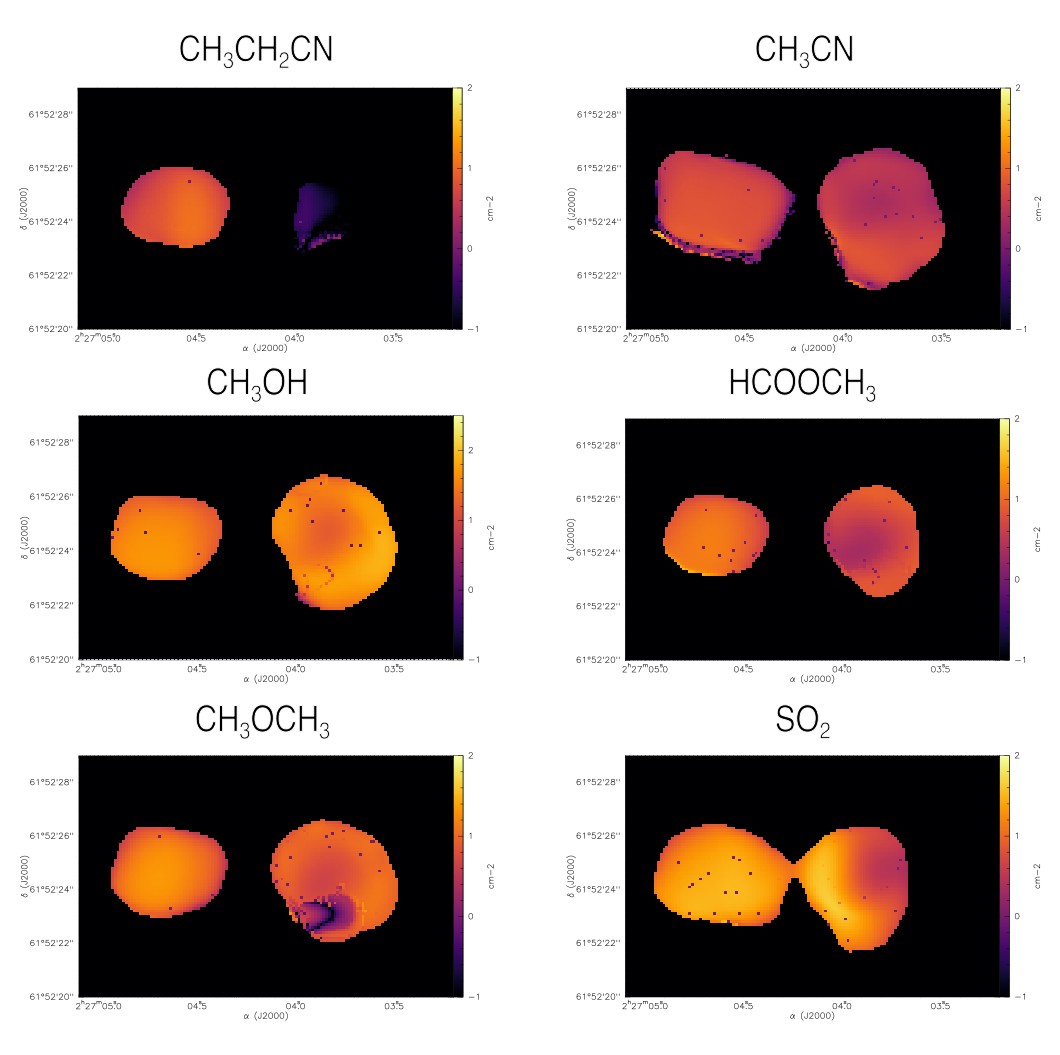}
  \caption{Maps describing the uncertainty of column density values for each molecule.
  \label{fig:uncertaintycoldens}}
\end{figure*}

\begin{figure*}[t!]
  \includegraphics[width=\textwidth]{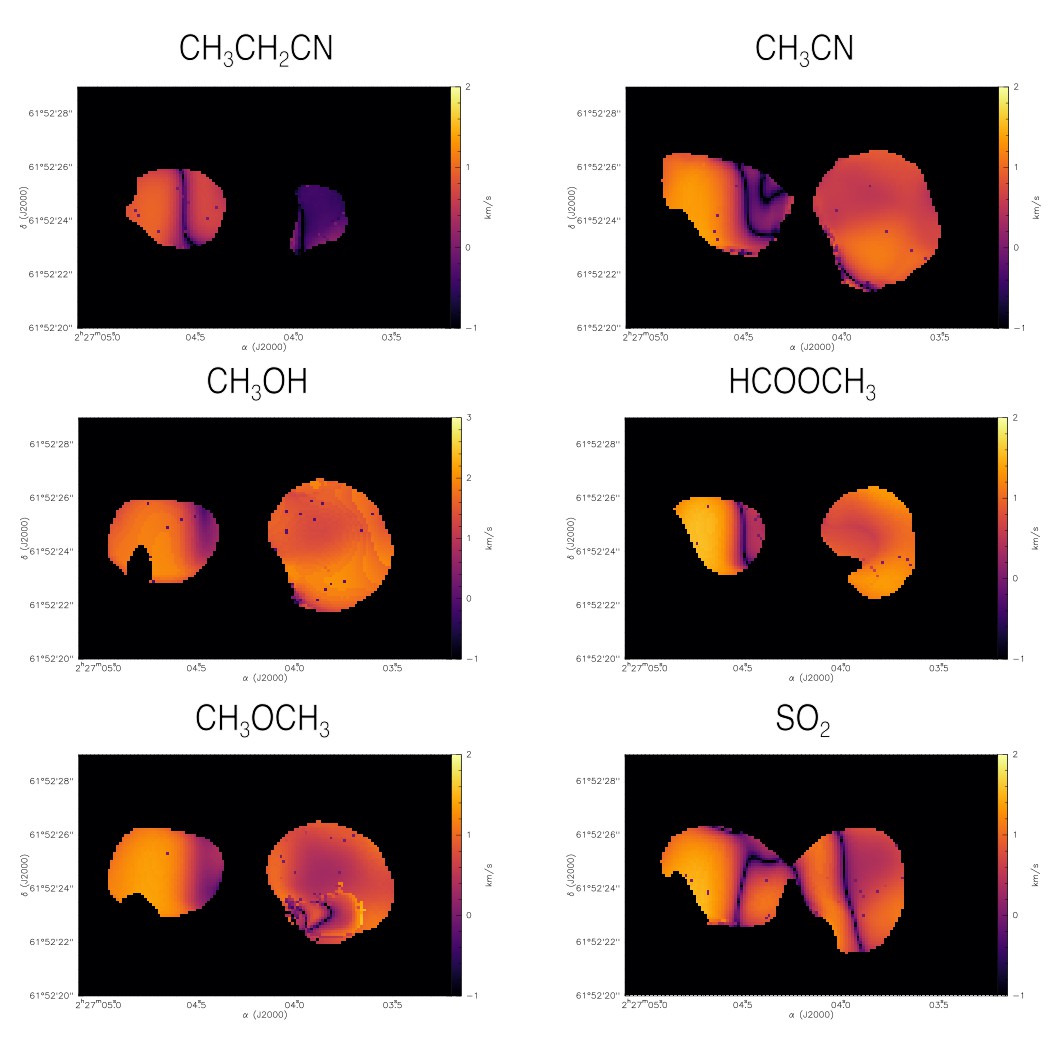}
  \caption{Maps describing the uncertainty of velocity shift values for each molecule.
  \label{fig:uncertaintyvel}}
\end{figure*}

This section provides uncertainty values for the physical parameters derived using GOBASIC. For each individual pixel, the derived value of temperature, column density, and velocity shift described in Figures \ref{fig:temp_maps}, \ref{fig:dens_maps}, and \ref{fig:vel_maps} was divided by the corresponding uncertainty. Figures \ref{fig:uncertaintytemp}, \ref{fig:uncertaintycoldens}, and \ref{fig:uncertaintyvel} plot these fractions spatially on a logarithmic scale.  These fractions represent two important physical values.  First, they are the inverse of the fractional uncertainty.  Second, they are the signal-to-noise ratio at each pixel.  The more negative the value in the plot, the higher the fractional uncertainty and the lower the signal-to-noise.

\end{appendix}

\bibliography{bib-doc}{}

\begin{thebibliography}{}
\expandafter\ifx\csname natexlab\endcsname\relax\def\natexlab#1{#1}\fi
\providecommand{\url}[1]{\href{#1}{#1}}
\providecommand{\dodoi}[1]{doi:~\href{http://doi.org/#1}{\nolinkurl{#1}}}
\providecommand{\doeprint}[1]{\href{http://ascl.net/#1}{\nolinkurl{http://ascl.net/#1}}}
\providecommand{\doarXiv}[1]{\href{https://arxiv.org/abs/#1}{\nolinkurl{https://arxiv.org/abs/#1}}}

\bibitem[{Ahmadi {et~al.}(2018)Ahmadi, Beuther, Mottram, Bosco, Linz, Henning,
  Winters, Kuiper, Pudritz, S{\'{a}}nchez-Monge, Keto, Beltran, Bontemps,
  Cesaroni, Csengeri, Feng, Galvan-Madrid, Johnston, Klaassen, Leurini,
  Longmore, Lumsden, Maud, Menten, Moscadelli, Motte, Palau, Peters, Ragan,
  Schilke, Urquhart, Wyrowski, \& Zinnecker}]{Ahmadi2018}
Ahmadi, A., Beuther, H., Mottram, J.~C., {et~al.} 2018, A\&A, 618

\bibitem[{Altwegg {et~al.}(2019)Altwegg, Balsiger, \& Fuselier}]{Altwegg2019}
Altwegg, K., Balsiger, H., \& Fuselier, S.~A. 2019, Annual Review of A\&A, 57,
  113

\bibitem[{{Astropy Collaboration} {et~al.}(2013){Astropy Collaboration},
  {Robitaille}, {Tollerud}, {Greenfield}, {Droettboom}, {Bray}, {Aldcroft},
  {Davis}, {Ginsburg}, {Price-Whelan}, {Kerzendorf}, {Conley}, {Crighton},
  {Barbary}, {Muna}, {Ferguson}, {Grollier}, {Parikh}, {Nair}, {Unther},
  {Deil}, {Woillez}, {Conseil}, {Kramer}, {Turner}, {Singer}, {Fox}, {Weaver},
  {Zabalza}, {Edwards}, {Azalee Bostroem}, {Burke}, {Casey}, {Crawford},
  {Dencheva}, {Ely}, {Jenness}, {Labrie}, {Lim}, {Pierfederici}, {Pontzen},
  {Ptak}, {Refsdal}, {Servillat}, \& {Streicher}}]{astropy:2013}
{Astropy Collaboration}, {Robitaille}, T.~P., {Tollerud}, E.~J., {et~al.} 2013,
  \aap, 558, A33

\bibitem[{{Astropy Collaboration} {et~al.}(2018){Astropy Collaboration},
  {Price-Whelan}, {Sip{\H{o}}cz}, {G{\"u}nther}, {Lim}, {Crawford}, {Conseil},
  {Shupe}, {Craig}, {Dencheva}, {Ginsburg}, {Vand erPlas}, {Bradley},
  {P{\'e}rez-Su{\'a}rez}, {de Val-Borro}, {Aldcroft}, {Cruz}, {Robitaille},
  {Tollerud}, {Ardelean}, {Babej}, {Bach}, {Bachetti}, {Bakanov}, {Bamford},
  {Barentsen}, {Barmby}, {Baumbach}, {Berry}, {Biscani}, {Boquien}, {Bostroem},
  {Bouma}, {Brammer}, {Bray}, {Breytenbach}, {Buddelmeijer}, {Burke},
  {Calderone}, {Cano Rodr{\'\i}guez}, {Cara}, {Cardoso}, {Cheedella}, {Copin},
  {Corrales}, {Crichton}, {D'Avella}, {Deil}, {Depagne}, {Dietrich}, {Donath},
  {Droettboom}, {Earl}, {Erben}, {Fabbro}, {Ferreira}, {Finethy}, {Fox},
  {Garrison}, {Gibbons}, {Goldstein}, {Gommers}, {Greco}, {Greenfield},
  {Groener}, {Grollier}, {Hagen}, {Hirst}, {Homeier}, {Horton}, {Hosseinzadeh},
  {Hu}, {Hunkeler}, {Ivezi{\'c}}, {Jain}, {Jenness}, {Kanarek}, {Kendrew},
  {Kern}, {Kerzendorf}, {Khvalko}, {King}, {Kirkby}, {Kulkarni}, {Kumar},
  {Lee}, {Lenz}, {Littlefair}, {Ma}, {Macleod}, {Mastropietro}, {McCully},
  {Montagnac}, {Morris}, {Mueller}, {Mumford}, {Muna}, {Murphy}, {Nelson},
  {Nguyen}, {Ninan}, {N{\"o}the}, {Ogaz}, {Oh}, {Parejko}, {Parley}, {Pascual},
  {Patil}, {Patil}, {Plunkett}, {Prochaska}, {Rastogi}, {Reddy Janga},
  {Sabater}, {Sakurikar}, {Seifert}, {Sherbert}, {Sherwood-Taylor}, {Shih},
  {Sick}, {Silbiger}, {Singanamalla}, {Singer}, {Sladen}, {Sooley},
  {Sornarajah}, {Streicher}, {Teuben}, {Thomas}, {Tremblay}, {Turner},
  {Terr{\'o}n}, {van Kerkwijk}, {de la Vega}, {Watkins}, {Weaver}, {Whitmore},
  {Woillez}, {Zabalza}, \& {Astropy Contributors}}]{astropy:2018}
{Astropy Collaboration}, {Price-Whelan}, A.~M., {Sip{\H{o}}cz}, B.~M., {et~al.}
  2018, \aj, 156, 123

\bibitem[{{Astropy Collaboration} {et~al.}(2022){Astropy Collaboration},
  {Price-Whelan}, {Lim}, {Earl}, {Starkman}, {Bradley}, {Shupe}, {Patil},
  {Corrales}, {Brasseur}, {N{"o}the}, {Donath}, {Tollerud}, {Morris},
  {Ginsburg}, {Vaher}, {Weaver}, {Tocknell}, {Jamieson}, {van Kerkwijk},
  {Robitaille}, {Merry}, {Bachetti}, {G{"u}nther}, {Aldcroft},
  {Alvarado-Montes}, {Archibald}, {B{'o}di}, {Bapat}, {Barentsen}, {Baz{'a}n},
  {Biswas}, {Boquien}, {Burke}, {Cara}, {Cara}, {Conroy}, {Conseil}, {Craig},
  {Cross}, {Cruz}, {D'Eugenio}, {Dencheva}, {Devillepoix}, {Dietrich},
  {Eigenbrot}, {Erben}, {Ferreira}, {Foreman-Mackey}, {Fox}, {Freij}, {Garg},
  {Geda}, {Glattly}, {Gondhalekar}, {Gordon}, {Grant}, {Greenfield}, {Groener},
  {Guest}, {Gurovich}, {Handberg}, {Hart}, {Hatfield-Dodds}, {Homeier},
  {Hosseinzadeh}, {Jenness}, {Jones}, {Joseph}, {Kalmbach}, {Karamehmetoglu},
  {Ka{l}uszy{'n}ski}, {Kelley}, {Kern}, {Kerzendorf}, {Koch}, {Kulumani},
  {Lee}, {Ly}, {Ma}, {MacBride}, {Maljaars}, {Muna}, {Murphy}, {Norman},
  {O'Steen}, {Oman}, {Pacifici}, {Pascual}, {Pascual-Granado}, {Patil},
  {Perren}, {Pickering}, {Rastogi}, {Roulston}, {Ryan}, {Rykoff}, {Sabater},
  {Sakurikar}, {Salgado}, {Sanghi}, {Saunders}, {Savchenko}, {Schwardt},
  {Seifert-Eckert}, {Shih}, {Jain}, {Shukla}, {Sick}, {Simpson},
  {Singanamalla}, {Singer}, {Singhal}, {Sinha}, {Sip{H{o}}cz}, {Spitler},
  {Stansby}, {Streicher}, {{{S}}umak}, {Swinbank}, {Taranu}, {Tewary},
  {Tremblay}, {Val-Borro}, {Van Kooten}, {Vasovi{'c}}, {Verma}, {de Miranda
  Cardoso}, {Williams}, {Wilson}, {Winkel}, {Wood-Vasey}, {Xue}, {Yoachim},
  {Zhang}, {Zonca}, \& {Astropy Project Contributors}}]{astropy:2022}
{Astropy Collaboration}, {Price-Whelan}, A.~M., {Lim}, P.~L., {et~al.} 2022,
  \apj, 935, 167

\bibitem[{Bisschop {et~al.}(2007)Bisschop, J{\o}rgensen, {van Dishoeck}, \& {de
  Wachter}}]{Bisschop2007}
Bisschop, S.~E., J{\o}rgensen, J.~K., {van Dishoeck}, E.~F., \& {de Wachter},
  E. B.~M. 2007, A\&A, 465, 913

\bibitem[{Blake \& {van Dishoeck}(1998)}]{Blake1998}
Blake, G.~A., \& {van Dishoeck}, E.~F. 1998, Annual Review of A\&A, 36, 317

\bibitem[{Brouillet {et~al.}(2022)Brouillet, Despois, Molet, Nony, Motte,
  Gusdorf, Louvet, Bontemps, Herpin, Bonfand, Csengeri, Ginsburg, Cunningham,
  Galván-Madrid, Maud, Busquet, Bronfman, Fernández-López, Jeff, Lefloch,
  Pouteau, Sanhueza, Stutz, \& Valeille-Manet}]{Brouillet2022}
Brouillet, N., Despois, D., Molet, J., {et~al.} 2022, A\&A, 665, A140

\bibitem[{Charnley(1997)}]{Charnley1997}
Charnley, S.~B. 1997, ApJ, 481, 396

\bibitem[{Chen {et~al.}(2006)Chen, Welch, Wilner, \& Sutton}]{Chen2006}
Chen, H., Welch, W.~J., Wilner, D.~J., \& Sutton, E.~C. 2006, ApJ, 639, 975

\bibitem[{Chyba \& Sagan(1992)}]{Chyba1992}
Chyba, C., \& Sagan, C. 1992, Nature, 355, 125

\bibitem[{Chyba {et~al.}(1990)Chyba, Thomas, Brookshaw, \& Sagan}]{Chyba1990}
Chyba, C.~F., Thomas, P.~J., Brookshaw, L., \& Sagan, C. 1990, Science, 249,
  366

\bibitem[{Garrod \& Herbst(2006)}]{Garrod2006}
Garrod, R., \& Herbst, E. 2006, A\&A, 457, 927

\bibitem[{Garrod {et~al.}(2022)Garrod, {Mihwa, J.}, {Matis, K.}, {Jones, D.},
  \& {Willis, E.}and {Herbst, E.}}]{Garrod2022}
Garrod, R., {Mihwa, J.}, {Matis, K.}, {Jones, D.}, \& {Willis, E.}and {Herbst,
  E.} 2022, ApJS, 249, 26

\bibitem[{Garrod {et~al.}(2008)Garrod, {Widicus Weaver, S.L.}, \& {Herbst,
  E.}}]{Garrod2008}
Garrod, R., {Widicus Weaver, S.L.}, \& {Herbst, E.} 2008, ApJ, 682, 283

\bibitem[{Gorai {et~al.}(2021)Gorai, Das, Shimonishi, Sahu, Kumar~Mondal, Bhat,
  \& Chakrabarti}]{Gorai2021}
Gorai, P., Das, A., Shimonishi, T., {et~al.} 2021, ApJ, 907, 108

\bibitem[{Hartogh {et~al.}(2011)Hartogh, Lis, Bockel{\'{e}}e-Morvan, {De
  Val-Borro}, Biver, K{\"{u}}ppers, Emprechtinger, Bergin, Crovisier, Rengel,
  Moreno, Szutowicz, \& Blake}]{Hartogh2011}
Hartogh, P., Lis, D.~C., Bockel{\'{e}}e-Morvan, D., {et~al.} 2011, Nature, 478,
  218

\bibitem[{Hatchell {et~al.}(1998)Hatchell, Thompson, Millar, \&
  Macdonald}]{Hatchell1998}
Hatchell, J., Thompson, M.~A., Millar, T.~J., \& Macdonald, G.~H. 1998, A\&A,
  338, 713

\bibitem[{Helmich {et~al.}(1994)Helmich, Jansen, de~Graauw, Groesbeck, \& van
  Dishoeck}]{Helmich1994}
Helmich, F., Jansen, D., de~Graauw, T., Groesbeck, T., \& van Dishoeck, E.
  1994, Astronomical Society of the Pacific, 283, 626

\bibitem[{Helmich \& van Dischoeck(1997)}]{Helmich1997}
Helmich, F., \& van Dischoeck, E. 1997, A\&AS, 124

\bibitem[{Herbst \& {van Dishoeck}(2009)}]{Herbst2009}
Herbst, E., \& {van Dishoeck}, E.~F. 2009, Annual Review of A\&A, 47, 427

\bibitem[{J{\o}rgensen {et~al.}(2012)J{\o}rgensen, Favre, Bisschop, Bourke,
  {van Dishoeck}, \& Schmalzl}]{Jorgensen2012}
J{\o}rgensen, J.~K., Favre, C., Bisschop, S.~E., {et~al.} 2012, Astrophysical
  Journal Letters, 757

\bibitem[{J{\o}rgensen {et~al.}(2004)J{\o}rgensen, Hogerheijde, Blake, {van
  Dishoeck}, Mundy, \& Sch{\"{o}}ier}]{Jorgensen2004}
J{\o}rgensen, J.~K., Hogerheijde, M.~R., Blake, G.~A., {et~al.} 2004, A\&A,
  415, 1021

\bibitem[{Keto {et~al.}(1995)Keto, Welch, Reid, \& Ho}]{Keto1995}
Keto, E.~R., Welch, W.~J., Reid, M.~J., \& Ho, P. T.~P. 1995, ApJ, 444, 765

\bibitem[{Laas {et~al.}(2011)Laas, Garrod, Herbst, \& Weaver}]{Laas2011}
Laas, J.~C., Garrod, Herbst, E., \& Weaver, S.~L. 2011, Astrophysical Journal,
  728

\bibitem[{Law {et~al.}(2021)Law, Zhang, {\"{O}}berg, Galv{\'{a}}n-Madrid, Keto,
  Liu, \& Ho}]{Law2021}
Law, C.~J., Zhang, Q., {\"{O}}berg, K.~I., {et~al.} 2021, ApJ, 909, 214

\bibitem[{Ligterink {et~al.}(2022)Ligterink, Ahmadi, Luitel, Coutens, Calcutt,
  Tychoniec, Linnartz, J{\o}rgensen, Garrod, \& Bouwman}]{Ligterink2022}
Ligterink, N.~F., Ahmadi, A., Luitel, B., {et~al.} 2022, ACS Earth and Space
  Chemistry, 6, 455

\bibitem[{Loren \& de~Mundy(1984)}]{Loren1984}
Loren, R.~B., \& de~Mundy, L.~G. 1984, ApJ, 286, 232

\bibitem[{Millar {et~al.}(1997)Millar, Macdonald, \& Gibb}]{Millar1997}
Millar, T.~J., Macdonald, G.~H., \& Gibb, A.~G. 1997, A\&A, 325, 1163

\bibitem[{Minh {et~al.}(2010)Minh, Su, Chen, Liu, Yan, \& Kim}]{Minh2010}
Minh, Y.~C., Su, Y.-N., Chen, H.-R., {et~al.} 2010, ApJ, 723, 1231

\bibitem[{M\"uller {et~al.}(2005)M\"uller, Schlöder, Stutzki, \&
  Winnewisser}]{Muller2005}
M\"uller, H., Schlöder, F., Stutzki, J., \& Winnewisser, G. 2005, Journal of
  Molecular Structure, 742, 215

\bibitem[{M\"uller {et~al.}(2001)M\"uller, {Thorwirth, S.}, {Roth, D. A.}, \&
  {Winnewisser, G.}}]{Muller2001}
M\"uller, H., {Thorwirth, S.}, {Roth, D. A.}, \& {Winnewisser, G.} 2001, A\&A,
  370, L49

\bibitem[{{Munoz Caro} {et~al.}(2002){Munoz Caro}, Meierhenrich, Schutte,
  Barbier, {Arcones Segovia}, Rosenbauer, Thiemann, Brack, \&
  Greenberg}]{MunozCaro2002}
{Munoz Caro}, G.~M., Meierhenrich, U., Schutte, W., {et~al.} 2002, Nature, 416,
  403

\bibitem[{{\"{O}}berg {et~al.}(2011){\"{O}}berg, Boogert, Pontoppidan, {Van Den
  Broek}, {van Dishoeck}, Bottinelli, Blake, \& Evans}]{Oberg2011}
{\"{O}}berg, K.~I., Boogert, A.~C., Pontoppidan, K.~M., {et~al.} 2011,
  Astrophysical Journal, 740

\bibitem[{Or{\'{o}}(1961)}]{Oro1961}
Or{\'{o}}, J. 1961, Nature, 190, 389

\bibitem[{Pickett {et~al.}(1998)Pickett, Poynter, Cohen, Delitsky, Pearson, \&
  M\"uller}]{Picket1998}
Pickett, H., Poynter, R., Cohen, E., {et~al.} 1998, Journal of Quantitative
  Spectroscopy and Radiative Transfer, 60, 883

\bibitem[{Qin {et~al.}(2015)Qin, {Schilke, P.}, {Wu, J.}, {Wu, Y.}, {Liu, T.},
  {Liu, Y.}, \& {S{\'{a}}nchez-Monge, {\'{A}}.}}]{Qin2015}
Qin, S.~L., {Schilke, P.}, {Wu, J.}, {et~al.} 2015, Astrophysical Journal, 803,
  39

\bibitem[{Rad {et~al.}(2016)Rad, Zou, Sanders, \& {Widicus Weaver}}]{Rad2016}
Rad, M.~L., Zou, L., Sanders, J.~L., \& {Widicus Weaver}, S.~L. 2016, A\&A,
  585, 1

\bibitem[{Rivera-Ingraham {et~al.}(2013)Rivera-Ingraham, Martin, Polychroni,
  Motte, Schneider, Bontemps, Hennemann, Men'Shchikov, Luong, Andr{\'{e}},
  Arzoumanian, Bernard, {Di Francesco}, Elia, Fallscheer, Hill, Li, Minier,
  Pezzuto, Roy, Rygl, Sadavoy, Spinoglio, White, \&
  Wilson}]{Rivera-Ingraham2013}
Rivera-Ingraham, A., Martin, P.~G., Polychroni, D., {et~al.} 2013,
  Astrophysical Journal, 766

\bibitem[{Rivilla {et~al.}(2020)Rivilla, Mart{\'{i}}n-Pintado,
  Jim{\'{e}}nez-Serra, Mart{\'{i}}n, Rodr{\'{i}}guez-Almeida, Requena-Torres,
  Rico-Villas, Zeng, \& Briones}]{Rivilla2020}
Rivilla, V.~M., Mart{\'{i}}n-Pintado, J., Jim{\'{e}}nez-Serra, I., {et~al.}
  2020, ApJ, 899, L28

\bibitem[{Stecklum {et~al.}(2002)Stecklum, Brandl, Henning, Pascucci, Hayward,
  \& Wilson}]{Stecklum2002}
Stecklum, B., Brandl, B., Henning, T., {et~al.} 2002, A\&A, 392, 1025

\bibitem[{Thompson {et~al.}(2023)Thompson, Giese, Lis, \&
  Widicus~Weaver}]{Thompson2023}
Thompson, W.~E., Giese, M.~M., Lis, D.~C., \& Widicus~Weaver, S.~L. 2023, ApJ,
  952, 50

\bibitem[{Tobin {et~al.}(2011)Tobin, Hartmann, Chiang, Looney, Bergin,
  Chandler, Masqu{\'{e}}, Maret, \& Heitsch}]{Tobin2011}
Tobin, J.~J., Hartmann, L., Chiang, H.~F., {et~al.} 2011, Astrophysical
  Journal, 740

\bibitem[{Tychoniec {et~al.}(2021)Tychoniec, {van Dishoeck}, {Van 'T Hoff},
  {Van Gelder}, Tabone, Chen, Harsono, Hull, Hogerheijde, Murillo, \&
  Tobin}]{Tycho2021}
Tychoniec, {\L}., {van Dishoeck}, E.~F., {Van 'T Hoff}, M.~L., {et~al.} 2021,
  A\&A, 655

\bibitem[{{van Dishoeck}(2006)}]{VanDishoeck2006}
{van Dishoeck}, E.~F. 2006, Proceedings of the National Academy of Sciences of
  the United States of America, 103, 12249

\bibitem[{Wakelam {et~al.}(2004)Wakelam, Caselli, Ceccarelli, Herbst, \&
  Castets}]{Wakelam2004}
Wakelam, V., Caselli, P., Ceccarelli, C., Herbst, E., \& Castets, A. 2004,
  A\&A, 422, 159

\bibitem[{{Widicus Weaver} \& Friedel(2012)}]{WidicusWeaver2012}
{Widicus Weaver}, S.~L., \& Friedel, D.~N. 2012, Astrophysical Journal,
  Supplement Series, 201

\bibitem[{{Widicus Weaver} {et~al.}(2017){Widicus Weaver}, Laas, Zou, Kroll,
  Rad, Hays, Sanders, Lis, Cross, Wehres, McGuire, \&
  Sumner}]{WidicusWeaver2017}
{Widicus Weaver}, S.~L., Laas, J.~C., Zou, L., {et~al.} 2017, ApJS, 232, 3

\bibitem[{Wilner {et~al.}(1999)Wilner, Reid, \& Menten}]{Wilner1999}
Wilner, D.~J., Reid, M.~J., \& Menten, K.~M. 1999, ApJ, 513, 775

\bibitem[{Wilner {et~al.}(1995)Wilner, Welch, \& Forster}]{Wilner1995}
Wilner, D.~J., Welch, W.~J., \& Forster, J.~R. 1995, ApJ, 449, 73

\bibitem[{Wright {et~al.}(2022)Wright, Smith, Kroll, Shipman, \&
  Widicus~Weaver}]{Wright2022}
Wright, C.~J., Smith, R.~N., Kroll, J.~A., Shipman, S.~T., \& Widicus~Weaver,
  S.~L. 2022, ACS Earth and Space Chemistry, 6, 482

\bibitem[{Wynn-Williams {et~al.}(1972)Wynn-Williams, Becklin, \&
  Neugebauer}]{Wynn1972}
Wynn-Williams, C., Becklin, E., \& Neugebauer, G. 1972, Monthly Notices of the
  Royal Astronomical Society, 160, 1

\bibitem[{Wyrowski {et~al.}(1997)Wyrowski, Hofner, Schilke, Walmsley, Wilner,
  \& Wink}]{Wyrowski1997}
Wyrowski, F., Hofner, P., Schilke, P., {et~al.} 1997, A\&A, 320, 17

\bibitem[{Wyrowski {et~al.}(1999)Wyrowski, Schilke, Walmsley, \&
  Menten}]{Wyrowski1999}
Wyrowski, F., Schilke, P., Walmsley, C., \& Menten, K. 1999, ApJ, 31

\bibitem[{Zhang {et~al.}(1998)Zhang, Ho, \& Ohashi}]{Zhang1998}
Zhang, Q., Ho, P. T.~P., \& Ohashi, N. 1998, ApJ, 494, 636

\bibitem[{Zou \& Widicus~Weaver(2017)}]{Zou2017}
Zou, L., \& Widicus~Weaver, S.~L. 2017, ApJ, 849, 139

\end{thebibliography}
\bibliographystyle{aasjournal}

\end{document}